\documentstyle[psfig,twocolumn,prl,aps,amssymb]{revtex}
\begin{document}
\draft

\date{June 25, 2002} 
\title{Theoretical Search for Nested Quantum Hall Effect of  
Composite Fermions}
\author{Sudhansu S. Mandal and Jainendra K. Jain}
\address{Department of Physics, 104 Davey Laboratory, The Pennsylvania State 
University, University Park, Pennsylvania 16802}
\maketitle

\begin{abstract}

Almost all quantum Hall effect to date can be understood as {\em integral}  
quantum Hall effect of appropriate particles, namely electrons or composite fermions.   
This paper investigates theoretically the feasibility of nested states of 
composite fermions which would lead to a quantum Hall effect  
that cannot be understood as integral quantum Hall effect of composite fermions.
The weak residual interaction between composite fermions will play a crucial role in 
the establishment of  
such quantum Hall states by opening a gap in a partially filled composite-fermion 
level.  To treat the problem of interacting composite fermions, 
we develop a powerful method that allows us to obtain the low energy spectra 
at composite fermion fillings of $\nu^*=n+\bar \nu$ without making any assumption regarding 
the structure of composite fermions in the topmost partially filled level.
The method is exact aside from neglecting the composite-fermion Landau level mixing, and 
enables us to study rather large systems, for example, 24 particles at a total flux of 62 $hc/e$, 
for which the dimension of the lowest Landau level Hilbert space is $\sim 10^{17}$.  
We have investigated, for fully spin polarized composite fermions, 
several filling factors between 1/3 and 2/5
using this approach.  The results indicate that any possible incompressibility 
at these fractions is likely to have a fundamentally different origin than that 
considered earlier.

\end{abstract}


\section{Introduction}

If electrons did not interact, only the integral quantum Hall effect 
(IQHE) \cite{Klitzing} would occur in nature.
The discovery of the fractional quantum Hall effect (FQHE) \cite{Tsui}
signaled the existence of a new correlated state of matter,  
the essence of which lies in the formation of quantum particles called 
composite fermions, \cite{Jain} which are bound 
states of electrons and an even number of quantized vortices.
Electrons avoid each other most efficiently by turning into 
composite fermions, which, in turn, interact much more weakly than electrons.
For a large number of phenomena, it is a valid first approximation to neglect the
interaction between composite fermions altogether, indicating that 
the interaction between composite fermions can often be treated perturbatively and does not 
cause any phase transitions.

The fundamental property of composite fermions, which is responsible for the dramatic 
phenomenology of two-dimensional systems in high magnetic fields, is that as they move 
about, the Aharonov Bohm 
phase is partly canceled by the phase produced by the vortices tied to other composite 
fermions and as a result they experience an {\em effective} magnetic field $B^*$ given by
\begin{equation}
B^*=B-2p\rho\phi_0
\label{B}
\end{equation}
where $B$ is the external magnetic field, $\rho$ is the electron density (in two dimensions), the even
integer $2p$ is the vorticity of composite fermions (CF's), and $\phi_0=hc/e$. 
(The composite fermion with vorticity $2p$ is denoted by $^{2p}$CF or CF-$2p$.)
They form Landau-like levels in this reduced magnetic field, with their filling factor ($\nu^*$) related
to the electron filling factor ($\nu$) by
\begin{equation}
\nu=\frac{\nu^*}{2p\nu^*\pm 1}
\label{nu}
\end{equation}
The wave functions for composite fermions are given by \cite{Jain}
\begin{equation}
\Psi_{\nu}=P_{LLL} \prod_{j<k}(z_j-z_k)^{2p} \Phi_{\nu^*}
\label{wf}
\end{equation}
where $z_j=x_j-iy_j$ denotes the position of the $j$th particle, 
$\Phi_{\nu^*}$ is an
antisymmetric wave function for fermions at the effective
filling $\nu^*$, and $P_{LLL}$ is the lowest Landau level (LLL) projection
operator.  The wave function $\Psi_{\nu}$ is of course the wave function of correlated 
electrons at $\nu$, but can also be interpreted as the wave function 
of composite fermions at filling factor $\nu^*$, because on the right hand side, 
each electron has $2p$ vortices bound to it through the multiplicative 
factor $\prod_{j<k}(z_j-z_k)^{2p}$, which converts each electron in 
into a composite fermion.

Let us begin by neglecting the interaction between composite fermions.
In this case, a gap opens up when the composite fermions 
fill an integral number of levels, i.e., when $\nu^*=n$, which corresponds to 
electron filling factors given by
\begin{equation}
\nu=\frac{n}{2pn+1}
\label{f1}
\end{equation}
where $n$ is a positive or negative integer.
[The negative values of $n$ give $\nu=|n|/(2p|n|-1)$, and correspond to the situation 
in which the effective field $B^*$ is opposite to the external field $B$.] 
The model of non-interacting composite fermions thus predicts quantum Hall effect
(QHE) of electrons at 
these fractions, and also at  
\begin{equation}
\nu=1-\frac{n}{2pn+ 1}
\label{f2}
\end{equation}
due to particle hole symmetry in the lowest Landau level.
(The fractions in Eq.~\ref{f2} can be obtained by formulating the problem in 
terms of {\em holes} rather than electrons 
in the lowest Landau level, and then making composite fermions out of them.)
The QHE at these filling factors will be termed ``CF-$2p$ QHE."  CF-2 QHE and CF-4 QHE 
are routine, and CF-6 QHE and CF-8 QHE have also been 
observed \cite{Goldman,Pan}.  
These sequences exhaust most of the more than 50 observed fractions to date in the lowest Landau 
level.   The wave functions have been tested in detail for finite systems, and are 
very accurate, without involving any adjustable parameter\cite{Jain,Dev,JK,Book2}.

In short, most of the QHE can be understood as the integral QHE of appropriate fermions.
It ought to be noted that the QHE of higher order composite fermions 
can be viewed as a consequence of interactions between lower order composite 
fermions.  For example, the CF-4 QHE originates from interactions between CF-2's.
Nonetheless, the physics of all states at filling factors of Eqs.~(\ref{f1}) and (\ref{f2}) 
is correctly  described in terms of {\em non}-interacting composite fermions of the 
appropriate kind.

While the above fractions are obtained most immediately in the CF theory, 
it has been known since the very beginning of the composite fermion theory that other 
fractions are not ruled out.\cite{Jain,Jain2}
New QHE {\em between} two successive integral Hall states of composite fermions
will be called NQHE, where N stands for ``nested." 
A simple scenario for NQHE is as follows.
Consider electrons in the filling factor range
\begin{equation}
\frac{n+1}{2p(n+1)+1} > \nu > \frac{n}{2pn+1}
\label{eq2}
\end{equation}
which map into $^{2p}$CF's in the range 
\begin{equation}
n+1 > \nu^* > n\;, 
\end{equation}
containing 
$n$ levels fully occupied and the $(n+1)^{st}$ level partially occupied. 
It is a natural conjecture that inter-CF interaction can possibly cause gaps at 
\begin{equation}
\nu^*=n + \frac{\bar{n}}{2\bar{p}\bar{n}+1}
\end{equation}
which will produce new fractions between the familiar fractions $\frac{n}{2pn+1}$ and 
$\frac{n+1}{2p(n+1)+1}$.  Here, the $^{2p}$CF's in the topmost partially filled level 
capture $2\bar{p}$ vortices (as a result of the residual $^{2p}$CF-$^{2p}$CF interaction)
to turn into higher order composite fermions ($^{2p+2\bar{p}}$CF's) and condense
into $\bar{n}$ levels, thereby opening a gap and producing quantum Hall effect. 
This QHE state is a nested state with higher order composite fermions forming in the 
background of the original composite fermions.

Certain qualitative consequences of the above scenario are as follows.
The NQHE of composite fermions is analogous to QHE of electrons 
between two successive IQHE states of 
electrons, that is, to FQHE in {\em higher} electronic Landau levels.
It is known that FQHE is rare in second electronic Landau levels (i.e., in the 
range $4>\nu>2$, including spin degeneracy) and non-existent in third or higher 
Landau levels; by analogy, NQHE is expected to be rare.
Extending the analogy further, the 
strongest NQHE states for fully spin polarized composite fermions
are expected to be at $\nu^*=1+1/3$ and $\nu^*=1+2/3$, which 
correspond to electron filling factors $\nu=4/11$ and $\nu=5/13$; these are expected to
be the strongest nested CF states \cite{Jain,Jain2}.

Recently, Pan {\em et al.} \cite {Pan2} have reported observation of 4/11 and 5/13, 
heralding a new generation of quantum Hall effect that cannot be understood as the  
IQHE of composite fermions.  This discovery has given fresh impetus to the issue of 
NQHE, especially because it is possible that more NQHE states
will be observed in the future as the sample quality improves and 
the temperature drops, reminiscent of how the FQHE appeared on top of the IQHE.
What other nested states are possible?  What is the true nature of these states?  
Such questions have motivated us to investigate this topic further.

Even though the theoretical scenario described above is plausible, and indeed natural,
there is no {\em a priori} guarantee that it 
actually occurs, and it is important to carry out quantitative tests to ascertain 
its applicability to the real world.  It is worth recalling here that not all 
fractions that can occur {\em in principle} do really occur in nature.  
For example, FQHE can occur for some hypothetical interaction at very small filling factors, but 
is preempted by Wigner crystal 
for the Coulomb interaction, and in higher Landau levels, the 
FQHE is often believed to be unstable to charge-density-wave type state.
Therefore, it would require further theoretical work before one could claim with any degree 
of confidence that NQHE based on the physics described above is actually possible.

The quantitative theoretical investigations so far suggest that there should be 
{\em no} FQHE at fractions other than 
those in Eqs.~(\ref{f1}) and (\ref{f2}) 
for fully spin-polarized electrons.  An exact diagonalization study \cite{Quinn} 
of $N=8$ electrons at 4/11 finds a non-uniform ground state, which 
is an evidence against FQHE liquid.
In another approach \cite{Quinn,LSJ}, a model has been developed for the 
effective interaction between composite fermions in the second level.  In Ref.~\onlinecite{Quinn} 
it was argued that no new FQHE is obtained \cite{Quinn}.  Ref.~\onlinecite{LSJ} 
found that the form of this interaction favors a ``bubble crystal" of composite fermions
at filling factors like 4/11 rather than the quantum 
Hall effect.  These studies indicate that the effective interaction between composite fermions is 
not sufficiently strongly repulsive to stabilize FQHE at 4/11 and several other fractions. 
Both approaches have their problems. In exact diagonalization, the system is effectively 
small, with only $\sim 3$ composite fermions in the second 
level, and one may question if the conclusion based on this system will hold up as $N$ increases.
The effective interaction model involves several assumptions, because of which the 
results are plausible but not conclusive.

In Ref.~\onlinecite{Park} showed that FQHE at 4/11 is very likely possible for a {\em partially} spin
polarized system or a spin unpolarized system.  It is not yet known for sure if the FQHE states at 
4/11 and 5/13 are partially or fully polarized in the experiments of Pan {\em et al.}, 
\cite{Pan2} but given the relatively high magnetic fields at which these states have been observed, 
it is quite possible that they are fully spin polarized, and 
therefore it is worth revisiting the issue of whether fully polarized incompressible states can 
be found theoretically at these filling fractions.  That is the prime motivation behind 
our present study.  In this work, we will look for FQHE at several filling
fractions in the range between 1/3 and 2/5.

We will study finite systems of $N$ particles at total flux $2Q$ (in units of the flux 
quantum $\phi_0=hc/e$), where the relation between $N$ and $Q$ for a given filling factor 
will be derived from the physics described above.  The most reliable theoretical method 
would be exact diagonalization.  However, because of the 
exponentially growing Hilbert space, exact diagonalization is possible 
only for very small systems, and very few 
filling factors.  For example, for $\nu=4/11$, exact diagonalization has been 
possible for $N=8$ electrons.\cite{warning}  The next 
system contains $N=12$ (in the spherical geometry, discussed below), for which the lowest 
Landau level Hilbert space contains a total of $8.6\times 10^7$ Slater
determinant basis states; the diagonalization of matrices of such sizes is 
beyond the reach of present day computers.  We have developed a powerful new method, namely 
diagonalization in low-energy composite-fermion basis (LECFB), which we believe gives  
accurate results.  It exploits the fact that the composite fermion theory allows us to directly 
identify low-energy states of the full Hilbert space, 
which lets us work within a sub-space much smaller than the  
full Hilbert space. For example, we have been able to obtain the low-energy spectrum for the 4/11 state 
with as many as 24 electrons, where the dimension of the full Hilbert space is $\sim 10^{17}$.

\begin{figure}
\centerline{\psfig{figure=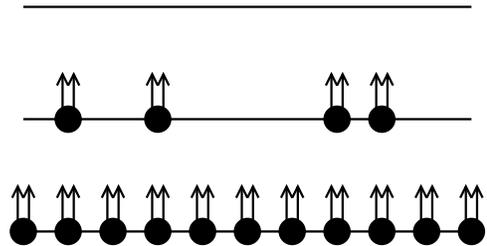,height=4.0in,angle=-90}}
\caption{The configurations included in our study.  Each composite fermions is depicted as 
an electron carrying two flux quanta.  The lowest composite fermion 
level is fully occupied and the second one 
is partially occupied.}
\label{fig_1}
\end{figure}

\begin{figure}
\centerline{\psfig{figure=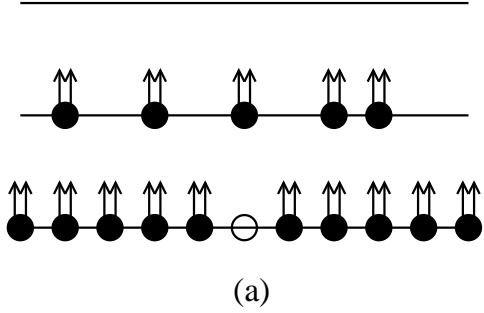,height=4.0in,angle=-90}}
\centerline{\psfig{figure=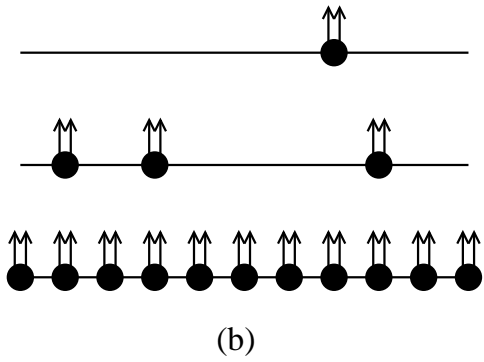,height=4.0in,angle=-90}}
\caption{The configurations neglected in our study.  These have higher ``kinetic" energy than 
the configurations shown in Fig.~\protect(\ref{fig_1}).}
\label{fig_2}
\end{figure}

A brief outline of the method is as follows.
Consider the filling factor range given in Eq.~(\ref{eq2}).
Given that the two ends of this range 
are well described as $n+1$ and $n$ filled levels of composite fermions, 
it is natural to expect that the low energy states at $\nu$ contain $n$ composite fermion levels 
completely occupied, and the remaining composite fermions in the $(n+1)^{st}$ level.
The configurations that involve promotion of composite fermions to higher kinetic energy levels
are expected to cost substantially higher energy and are neglected. 
In other words, we include all configurations 
of type shown in Fig.~(\ref{fig_1}) but neglect configurations
of the type shown in Fig.~(\ref{fig_2}).
This neglect of composite-fermion Landau level mixing is the only assumption in our approach,
which can be tested for consistency at the end of the calculation.  Other than that the method 
is exact. The most important point here is that we do not make any assumption with regard to 
the interaction between composite fermions, or their structure  
in the partially filled level.  In our calculation, we first construct all states of 
composite fermions at $\nu^*=n+\bar{\nu}$, with the lowest 
$n$ levels fully occupied and the next one with filling factor $\bar{\nu}$, and then 
diagonalize the Coulomb Hamiltonian within this sector to obtain the spectrum 
of low energy states.  The calculation is rather involved, requiring 
extensive Monte Carlo, but still numerically stable, and yields reliable results.

The present method is a significant advance over the variational method of Ref.~\onlinecite{LSJ}.
It is much more reliable because it eliminates several approximations 
made in that work.  In Ref.~\cite{LSJ}, the problem of composite fermions at $\nu^*=n+\bar{\nu}$ was 
mapped into that of fermions at $\bar{\nu}$.  Further, 
it was assumed that the interaction energy of the 
system of composite fermions in the partially filled level can be approximated by a sum of two-body 
interactions.  The interaction between composite fermions was obtained by keeping only two 
composite fermions in the $(n+1)^{st}$ level and integrating out the composite fermions of the 
lower filled levels.  In general, when there are many composite fermions in the $(n+1)^{st}$ level, 
the integrating out of the lower-level composite fermions will produce a quite complex interaction, 
with two-, three-, and $n$-body terms, because the full system is strongly correlated.  The hope 
was that the $n$-body terms do not cause any phase transitions, in which case the two-body terms 
will produce the correct state.  However, this approximation was untested. 
Finally, the method was a variational study, in which the energies of certain wave functions were 
compared to determine which had the lowest energy; there was of course no guarantee that this 
state would describe the true ground state.  In contrast, in 
the present study, we work with the full composite fermion system and do not assume anything 
about the form of the interaction between composite fermions or the nature of their ground state.

The following section contains a discussion of how we construct the low-energy composite fermion basis 
and how we carry out the Gram-Schmid orthogonalization and diagonalization.  The reader not interested in 
the technical details can skip this section and directly proceed to the subsequent section, wherein 
the results are given and their implications are discussed.

\section{Diagonalization in low-energy composite-fermion basis}

We first summarize certain relevant facts from the introduction.
The electron filling factor $\nu$ given by  
\begin{equation}
\nu=\frac{\nu^*}{2p\nu^*\pm 1}
\end{equation}
corresponds to composite fermion filling factor $\nu^*$. 
We will be interested below in the specific values of $\nu^*$ given by 
\begin{equation}
\nu^*=n+ \bar{\nu}=n+ \frac{\bar{n}}{2\bar{p}\bar{n}+1}
\end{equation}
Here, composite fermions fill $n$ levels completely and 
occupy $\bar{\nu}$ fraction of the $(n+1)^{st}$ level.
If the $^{2p}$CF's in the partially occupied level capture $2\bar{p}$ additional vortices and 
condense into $\bar{n}$ filled Landau level state, then a gap would open up and  
NQHE would be obtained.  The question is whether this mechanism really occurs in nature.

In the composite fermion theory, there are several possible approaches for constructing 
a low energy basis.  In this work, 
we construct a LECFB of {\em all} states at $\nu^*=n+\bar{\nu}$ 
in which $n$ $^{2p}$CF levels are fully occupied and the next level has filling $\bar{\nu}$. 
We do not make any approximation regarding the structure of $^{2p}$CF's in the partially filled 
level.  This basis is in principle straightforward:  we simply need to 
take all {\em electronic} states at $\nu^*=n+\bar{\nu}$ and 
attach $2p$ vortices to obtain the composite fermion states.  The method, however, is 
technically rather demanding, and the present section is devoted to it.

\subsection{Wave functions}

We will employ the spherical geometry in which
$N$ electrons move on the surface of a sphere under the
influence of a radial magnetic field $B$ created by a magnetic
monopole at the center.  The magnitude of
the $B$ field is given by $2Q\phi_0/ 4\pi R^2$ where
$\phi_0 = hc/ e$ is known as the flux quantum, $R$
is the radius of the sphere, and $Q$ is called the monopole
strength which should be either an integer or a half-integer
because of Dirac's quantization condition.

The interacting electron system at monopole strength $Q$ maps into a
system of weakly interacting composite fermions at an effective monopole
strength $q^*=Q-p(N-1)$.
The wave functions for interacting electrons at $Q$ are given by 
\begin{equation}
\Psi_{Q}=P_{LLL} \Phi_1^{2p} \Phi_{q^*}
\end{equation}
where $\Phi_1$ is the wave function for the fully occupied lowest Landau level, 
and $\Phi_{q^*}$ are antisymmetric wave functions at $q^*$.
$\Phi$ is in general a linear superposition of Slater-determinant
basis states made up of the monopole harmonics,
$Y_{q^*,s,m}$, given by \cite{Yang}:
\begin{eqnarray}
&&Y_{q^*,s,m}(\Omega_j)=N_{q^*sm} (-1)^{q^*+s-m}
e^{iq^*\phi_j} u_j^{q^*+m} v_j^{q^*-m}
\nonumber \\
&&\times\sum_{r=0}^{s}(-1)^r {{s \choose r}} {{ 2q^*+s
\choose q^*+s-m-r}}
(v_j^*v_j)^{s-r}(u_j^*u_j)^r\;\;,
\label{appndx_mh}
\end{eqnarray}
where
\begin{equation}
N_{q^* sm}=\left(\frac{(2q^*+2s+1)}{4\pi}\frac{(q^*+s-m)
!(q^*+s+m)!}{s!(2q^*+s)!}
\right) ^{1/2}\;\;.
\end{equation}
$s$=0,1,2,... is the Landau level (LL) index, to be differentiated from $n$, the number
of filled Landau levels.  $\Omega_j$ represents the angular coordinates
$\theta_j$ and $\phi_j$ of the $j$th electron, and
\begin{equation}
u_j \equiv \cos(\theta_j/2)\exp(-i\phi_j/2)
\end{equation}
\begin{equation}
v_j \equiv \sin(\theta_j/2)\exp(i\phi_j/2)\;.
\end{equation}

It was shown in Ref.~\onlinecite{JK} that 
$\Psi_Q$, the wave function of interacting
electrons at $Q= q^*+p(N-1)$, is obtained from $\Phi_{q^*}$ by replacing
$Y_{q^*,s,m}$ by $Y^{CF}_{q^*,s,m}$, defined as:
\begin{eqnarray}
&&Y_{q^*,s,m}^{CF}(\Omega_j) =
N_{q^*sm} (-1)^{q^*+s-m} \frac{(2Q+1)!}{(2Q+s+1)!}
u_j^{q^*+m}   v_j^{q^*-m}
\nonumber \\
&& \times\sum_{r=0}^{s}(-1)^r {{s \choose r}}
{{ 2q^*+s \choose q^*+s-m-r}}
\; u_j^r \;    v_j^{s-r}
\nonumber \\
&&
\times\; \left[ \left({\frac{\partial}{\partial u_j}} \right)^r
\;\left({\frac{\partial}{\partial v_j}}\right)^{s-r} J_j^p\right]\;
\end{eqnarray}

where
\begin{equation}
J_j=\prod_{k}^{'}
(u_j v_k-v_j u_k)
\end{equation}
Here the prime denotes the condition $k\neq j$.

\subsection{Low-energy composite-fermion basis}

For $\bar\nu=0$, that is at the special filling factors 
\begin{equation}
\nu=\frac{n}{2pn\pm 1}
\end{equation}
the wave function for the ground state 
has the simple form:
\begin{equation}
\Psi_{\frac{n}{2pn\pm 1}}^{GS}=P_{LLL}\Phi_1^{2p}\Phi_n
\end{equation}
where $\Phi_n$ is the Slater determinant wave function 
for $n$ filled Landau levels.  Since $\Phi_n$ is unique, the wave function $\Psi_{\frac{n}{2pn\pm 1}}$ 
also contains no adjustable parameters.  It has been tested in the past and was found to be remarkably 
accurate \cite{Dev,JK}.

Of interest in this work is the situation when $\bar \nu\neq 0$.
Here, we obtain the low energy spectrum through the following steps.

\begin{enumerate}

\item We perform exact diagonalization for $\bar N$ particles 
at filling factor $\bar{\nu}$ in the lowest Landau level to get all eigenstate. 
The form of the interaction used for the diagonalization 
is of no significance, because its only role is to produce basis wave functions, 
but we work with the Coulomb interaction.  
Let us denote these states by $\Phi^\alpha_{\bar \nu}$, where $\alpha$ labels different 
states.

\item We then promote each eigenstate to
the $(n+1)^{st}$ Landau level, and fill the lower $n$ Landau levels completely with additional 
particles to get the $N$ particle state at $\nu^*=n+\bar \nu$.
This gives {\em all} wave functions $\Phi^\alpha_{\nu^*}$ for which the lowest $n$ Landau levels
are fully occupied and the next Landau level has filling $\bar{\nu}$.
These states are denoted by $L_n \Phi^\alpha_{\bar \nu}$, where $L_n$ denotes addition of $n$ Landau levels.
It ought to be remembered that the operator $L_n$ changes the number of particles.

\item We then multiply each wave function by $\Phi_1^{2p}$ and carry out the lowest Landau 
level projection by the method discussed above.  This gives us correlated 
basis functions $\Psi^\alpha_\nu$ at $\nu$.  These steps are summarized as 

\begin{equation}
\bar \nu \Rightarrow \nu^*=n+\bar \nu \Rightarrow \nu=\frac{\nu^*}{2p\nu^*+1}
\end{equation} 

\begin{equation}
\Psi^\alpha_\nu= P_{LLL} \Phi_1^{2p} L_n \Phi^\alpha_{\bar \nu}
\end{equation}

The advantage of constructing basis states in this way is that it directly gives us basis
functions with well defined orbital angular momentum $L$, which is a good quantum number
in the spherical geometry.  We can work within each $L$ sector 
independently, because they are not coupled by the interaction.  
(Note that in the above equation, $\Phi^\alpha_{\bar \nu}$ is a wave function for $\bar N$ electrons 
only.)

\item The basis thus obtained is not orthogonal.  We obtain an orthogonal basis following the 
Gram-Schmid
procedure.  Some relevant details are outlined below.

\item We construct the Hamiltonian matrix.  

\item Diagonalization of the Hamiltonian gives the low energy eigenspectrum at $\nu$.

\end{enumerate}

If an incompressible state is obtained, the ground state wave function is 
likely to be well approximated by
\begin{equation}
P_{LLL} \Phi_1^{2p} L_n \Phi^{GS}_{\bar \nu}\approx 
P_{LLL} \Phi_1^{2p} L_n [P_{LLL}\Phi_1^{2\bar p}\Phi_{\bar n}]
\label{cfgs}
\end{equation} 
where $\Phi^{GS}_{\bar \nu} \approx P_{LLL}\Phi_1^{2\bar p}\Phi_{\bar n}$ is the wave function 
for the ground state for $\bar N$ particles at $\bar \nu=\frac{\bar n}{2\bar p \bar n +1}$.

\subsection{NQHE states on sphere}

Since our goal is to look the feasibility of NQHE 
described above, we do our calculations at those values of $Q$ where 
it is possible for $^{2p}$CF's in the partially filled level to condense into a standard 
type of FQHE state.  For that, we first need to 
determine the relation between $Q$ and $N$ for a given filling factor.  Clearly, in the
limit $N\rightarrow \infty$ we must have
\begin{equation}
\lim_{N\rightarrow \infty} \frac{N}{2Q}=\nu
\end{equation}
where $2Q$ is the number of flux quantum.  For finite systems, however, the ratio $N/2Q$ is not 
exactly equal to $\nu$, and the identification of what filling factor a finite system corresponds 
to requires additional theoretical input.

Let us first consider the incompressible states at 
filling factor $\bar{\nu}=\frac{\bar{n}}{2\bar{p}\bar{n}+1}$, the wave function for which
is given by $\Phi_1^{2\bar{p}} \Phi_{\bar{n}}$.  For these states, the 
relation between $Q$ and $N$ can be obtained by 
noting that the product of two monopole harmonics at $q'$ and $q''$ gives a monopole 
harmonic at $q'+q''$, i.e. the monopole strengths add. 
For reasons that will become clear below, let us denote the number of particles by 
$\bar{N}$ and the monopole strength by $\bar{q}$. 
The lowest LL projection is unimportant for the question of the
relationship between $\bar{N}$ and $\bar{q}$, as it does not alter $\bar{q}$.
In the spherical geometry, for monopole strength $q$ the degeneracy of the lowest Landau level is $2q+1$, 
for the next LL it is $2q+3$, and so on.  From that, it is clear that the $\bar{n}$ filled LL state is 
obtained for $q_{\bar n}$ given by
\begin{equation}
q_{\bar n}=\frac{\bar{N}-\bar{n}^2}{2\bar{n}}
\end{equation}
In particular, one filled LL is obtained at 
\begin{equation}
q_1=\frac{\bar{N}-1}{2}
\end{equation}
From the addition rule, the monopole strength of $\Phi_1^{2\bar{p}} \Phi_{\bar{n}}$ is given by
\begin{equation}
\bar{q}= 2\bar{p} q_1+ q_{\bar{n}} = 
    \left(\bar{p}+ \frac{1}{2\bar{n}}\right)\bar{N}-\left(\bar{p} + 
\frac{\bar{n}}{2}\right)
\end{equation}
It can be verified that $2\bar{q}/\bar{N}$ gives the filling factor $\bar \nu$ 
in the thermodynamic limit.

Now we ask what is the monopole strength $q^*$ for the state at 
$\nu^*=n+\bar{\nu}=n+\frac{\bar{n}}{2\bar{p}\bar{n}+ 1}$ for $N$ electrons.
Let us denote by $\bar{N}$ the number of fermions in the $(n+1)^{st}$ partially filled Landau level.
In order for $\bar{N}$ fermions 
to form the $\frac{\bar{n}}{2\bar{p}\bar{n}+ 1}$ state, the degeneracy in this level
must  satisfy:
\begin{equation}
2q^*+2n+1=2\bar{q}+1
\end{equation}
which implies that
\begin{equation}
q^*= \bar q - n
\end{equation}
The total number of fermions, including the fermions in the lowest $n$ fully occupied 
Landau levels is 
\begin{equation}
N=\bar{N}+2nq^*+n^2
\end{equation}
Eliminating $\bar{N}$, one gets the following relation between $N$ and $q^*$:
\begin{equation}
N=2q^*\left(n+\frac{\bar{n}}{2\bar{p}\bar{n}+1}\right)+\frac{\bar{p}
+\frac{1}{2}\bar{n}+n}{\bar{p}+\frac{1}{2\bar{n}}}+n^2
\end{equation}
The state at $\nu^*=n+\frac{\bar{n}}{2\bar{p}\bar{n}+1}$ is obtained for all vaules of ($N$,$q^*$)
related by the above equation, provided $N$ is an integer and $q^*$ is an integer or a half integer.
The state at $\nu=\frac{\nu^*}{2p\nu^*+1}$ in turn is obtained for all values of 
($N$,$Q$) with $Q=q^*+p(N-1)$.

We have considered in this work the following filling factors.  

\begin{itemize}

\item $\nu=\frac{4}{11}$:  
Here, $\nu^*=1+\frac{1}{3}$.  The parameters are: $p=\bar{p}=n=\bar{n}=1$, which 
gives

\begin{equation}
q^*=\frac{3N-8}{8}, \;\; Q=\frac{11N-16}{8}
\end{equation}

\item $\nu=\frac{5}{13}$:  Here, $\nu^*=1+\frac{2}{3}$.  The parameters are: $p=\bar{p}=n=1$ and 
$\bar{n}=-2$, which gives
\begin{equation}
q^*=\frac{3N-7}{10}, \;\; Q=\frac{13N-17}{10}
\end{equation}

\item
$\nu=\frac{7}{19}$:  Here, $\nu^*=1+\frac{2}{5}$.  The parameters are: $p=\bar{p}=n=1$ and 
$\bar{n}=2$, which gives
\begin{equation}
q^*=\frac{5N-17}{14}, \;\; Q=\frac{19N-31}{14}
\end{equation}

\item
$\nu=\frac{6}{17}$:  Here, $\nu^*=1+\frac{1}{5}$.  The parameters are: $p=n=\bar{n}=1$ and 
$\bar{p}=2$, which gives
\begin{equation}
q^*=\frac{5N-12}{12}, \;\; Q=\frac{17N-24}{12}
\end{equation}

\end{itemize}

Of course, only integer values are allowed for $N$, $2q^*$, and $2Q$.
The table \ref{tab1} explicitly lists the systems that we have studied below.

\subsection{Matrix elements}

The basis obtained in the way described above is not orthogonal for a given
angular momentum sector $L$.  We orthogonalize these
using the Gram-Schmid procedure 
and than diagonalize the Coulomb Hamiltonian in the orthogonal basis. 
These require the knowledge of matrix elements of the type 
$\big< \phi_l \vert \phi_m \big>$ and $\big< \phi_l \vert H \vert \phi_m \big>$, 
where $\phi_l$ and $\phi_m$
($l\neq m$) are unorthonormalized wave functions, which we obtain by Monte 
Carlo.  Since the Monte Carlo evaluation is most efficient when the integrand is positive definite,
we determine these using the equations \cite{Mandal,Bonesteel}
\begin{equation}
\big< \phi_l \vert \phi_m \big> = \frac{ \big< \phi_l + \phi_m \vert 
\phi_l +\phi_m \big>
      - \big< \phi_l \vert \phi_l \big> - \big< \phi_m \vert 
\phi_m \big> }{2}\;,
\end{equation}
and 
\begin{equation}
\big< \phi_l \vert V \vert \phi_m \big> = \frac{ \big< \phi_l + 
\phi_m \vert V \vert  \phi_l + \phi_m \big>
      - \big< \phi_l \vert V \vert  \phi_l \big> - 
 \big< \phi_m \vert V \vert \phi_m \big> }{2}\;,
\end{equation}
where $V$ denotes the Coulomb interaction Hamiltonian. 

\subsection{Orthogonalization}

We denote $\{u_l \}$ as the unorthogonal but normalized basis set 
obtained from the set $\{ \phi_l \}$, and define 
\begin{equation}
u_{ij} \equiv \big< u_i \vert u_j \big> = 
\frac{ \big< \phi_i \vert \phi_j \big>}{ \vert \phi_i \vert \vert
\phi_j \vert } 
\end{equation}
and
\begin{equation}
V_{ij} \equiv \big< u_i \vert V \vert  u_j \big> = 
\frac{ \big< \phi_i \vert V \vert \phi_j \big>}{ \vert \phi_i \vert \vert
\phi_j \vert } \, ; 
\end{equation}
these are real numbers.

We then follow the Gram-Schmid procedure to construct an orthogonal basis set
$\{ \psi_i \}$ in terms of the unorthogonal basis set $\{ u_i \}$.  
The former can be expressed as
\begin{eqnarray}
\psi_j &=& u_j - \sum_{i=1}^{j-1} \frac{1}{N_i^2}
    \big< \psi_i \vert u_j \big> \psi_i \, ,
 \nonumber \\
       &\equiv & u_j - \sum_{i=1}^{j-1}f_{ij} \psi_i \, ,
\label{GS}
\end{eqnarray}
with $N_i^2 = \big< \psi_i \vert \psi_i \big>$ 
and $f_{ij} = \big< \psi_i \vert u_j \big>/ N_i^2$.
While the above relation allows an iterative evaluation of $\{ \psi_i \}$, 
we find it more convenient to work with the following expressions:
\begin{equation}
N_i^2 = \big< \psi_i \vert \psi_i \big> = 1-\sum_{k=1}^{i-1}f_{ki}^2 N_k^2
\, \end{equation}
and
\begin{eqnarray}
f_{ij} &=& \frac{\big< \psi_i \vert u_j \big> }{ N_i^2} \nonumber \\
  &=& \frac{1}{N_i^2} \left[ \big< u_i - \sum_{k=1}^{i-1}
     f_{ki} \psi_k \vert u_j \big> \right] \nonumber \\
&=& \frac{1}{N_i^2} \left[ u_{ij} - \sum_{k=1}^{i-1}
    f_{ki}f_{kj}N_k^2 \right] \, , 
\end{eqnarray}
Eq.~(\ref{GS}) was used to obtain these expressions, as well as the fact
that $u_{ij}$ and $f_{ij}$ are real.
The Hamiltonian matrix elements 
\begin{equation}
V_{ij} = \frac{ \big< \psi_i \vert V \vert \psi_j \big> }{N_i N_j}
\end{equation}
in the basis set $\{ \psi_i \}$
are thus obtained as
\begin{eqnarray}
V_{ij}  &=& \frac{1}{N_iN_j} \left[ V_{ij} -
   \sum_{k=1}^{j-1} f_{kj}E_{ik}N_iN_k \right. \nonumber \\  
  & & \left.  -\sum_{l=1}^{i-1} f_{li} \left\{ E_{lj}N_lN_j
   + \sum_{k=1}^{j-1}f_{kj}E_{lk}N_lN_k \right\} \right] \, 
\end{eqnarray}
which are again obtained iteratively.
We finally diagonalize the matrix $V_{ij}$
to obtain the eigenvalue spectrum for a given $L$ sector.

\subsection{Computation}

The major computation time is spent in calculating the matrix elements
$\big< \phi_l \vert \phi_m\big>$ and $\big< \phi_l \vert  V \vert \phi_m\big>$
in the Monte Carlo method, because at each step we must evaluate  
$M$ Slater determinants, $M$ being the size of the LECFB.
To give an example, at $(N,Q)=(24,31)$ 
we have $M=1656$. To minimize the computational time, we sample all the 
states relative to the $L=0$ state, which is obtained from 
the lowest energy state at $\nu^*$. 
We typically perform six to ten Monte Carlo runs for each $(N,Q)$
with (0.6 - 1.0)$\times 10^6$  Monte Carlo iterations in each 
run. For larger systems, we use parallel computers.
(We divide all of the Monte Carlo steps into several configurations
with each configuration placed on a single node of a
Beowolf class PC cluster.  One node consists of a dual 1 GHZ
Pentium III Processor.  To obtain one data point, at a
particular $N$, we use as many as 30 nodes repeatedly until the
standard deviation in energies is sufficiently low to produce the desired
accuracy.)
Approximately 80 CPU days were needed for the calculation of the spectrum for $(N,Q)=(24,31)$. 
The computation time increases very quickly with $N$, approximately as $N^{12}$.

It is difficult to ascertain how the statistical error in each element of the matrix 
would propagate into the eigenvalues.  We instead obtain the spectrum in several 
different runs, and use the eigenvalues from those different runs to obtain the 
average and the standard deviation for each eigenvalue.

\section{Results and discussion}

\begin{figure}
\centerline{\psfig{figure=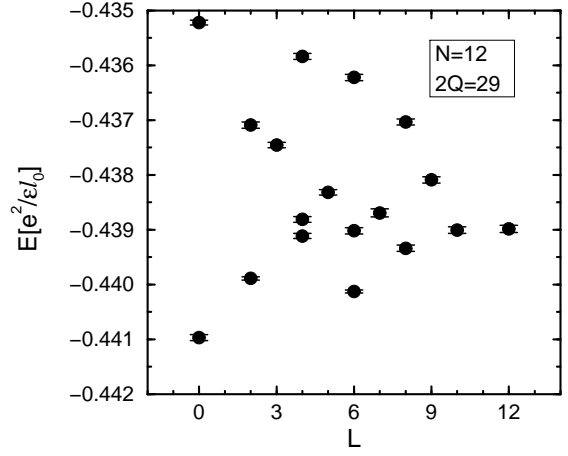,height=4.0in,angle=-90}}
\caption{The low energy spectrum of interacting composite fermions at $(N,2Q)=(12,29)$, 
identified with 4/11,
where $N$ is the number of composite fermions and $2Q$ is total flux penetrating the sample. 
The energies in this and subsequent figures 
are quoted in units of $e^2/\epsilon l_0$, where $\epsilon$ is the 
dielectric constant for the background material ($\epsilon \approx 13$ for GaAs) and 
$l_0=\sqrt{\hbar c/eB}$ is the magnetic length at the relevant filling factor (4/11 in this case). 
Spherical geometry is employed for the calculation, and 
$L$ is the orbital angular momentum of the state.}
\label{fig1}
\end{figure}

\begin{figure}
\centerline{\psfig{figure=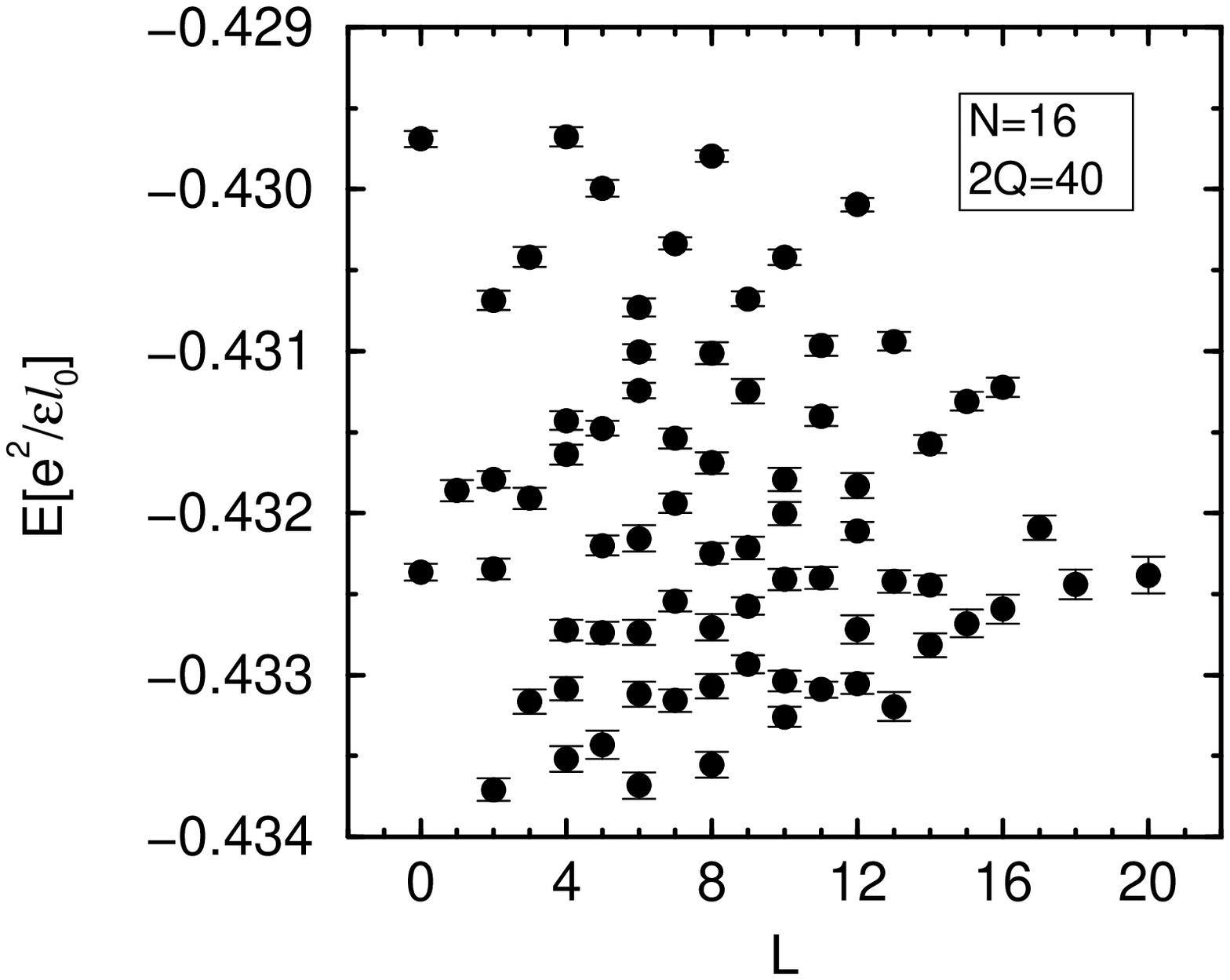,height=4.0in,angle=-90}}
\caption{The low energy spectrum of interacting composite fermions at $(N,2Q)=(16,40)$,
identified with 4/11.}
\label{fig2}
\end{figure}

\begin{figure}
\centerline{\psfig{figure=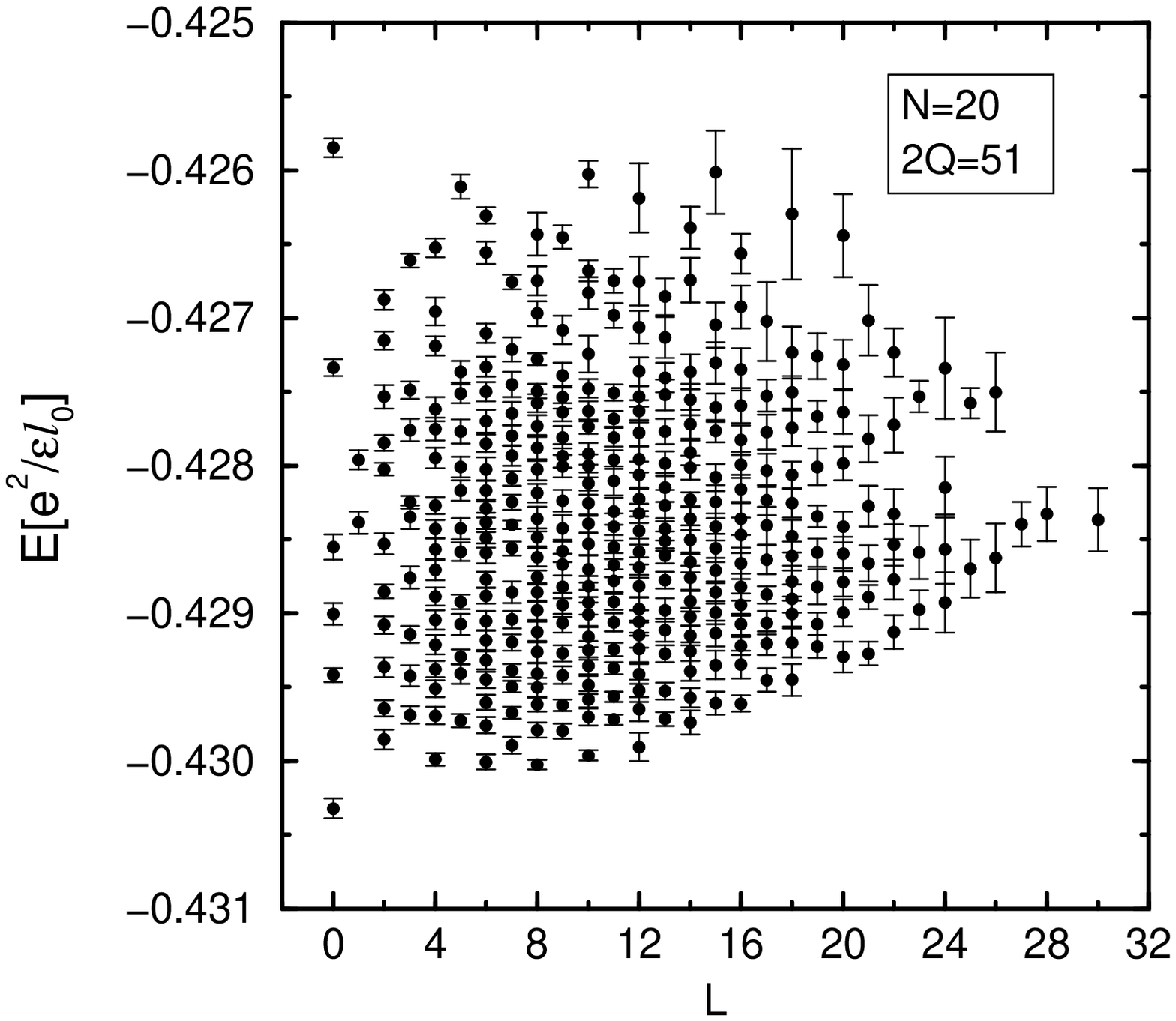,height=4.0in,angle=-90}}
\caption{The low energy spectrum of interacting composite fermions at $(N,2Q)=(20,51)$,
identified with 4/11.}
\label{fig3}
\end{figure}

\begin{figure}
\centerline{\psfig{figure=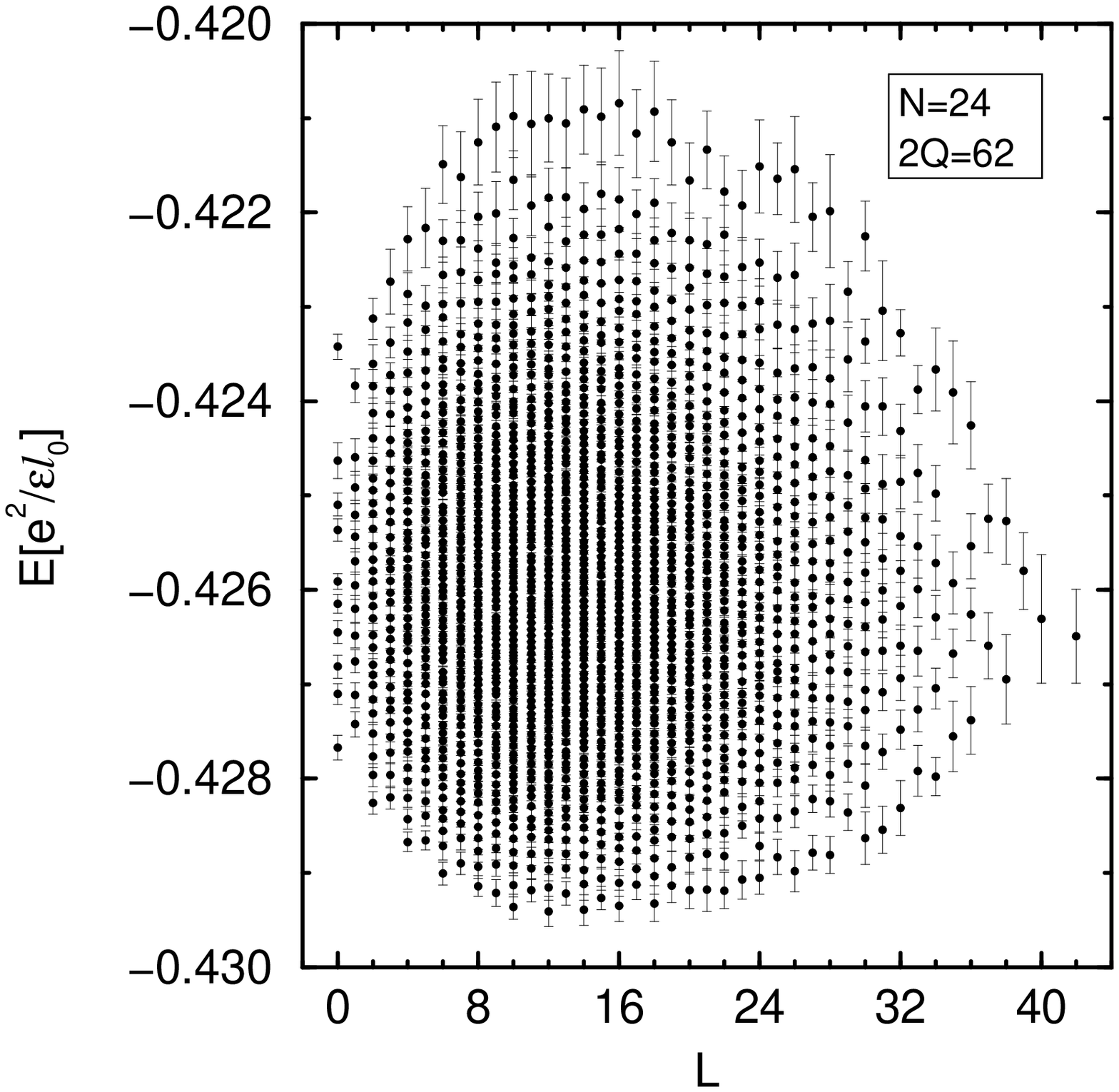,height=4.0in,angle=-90}}
\caption{The low energy spectrum of interacting composite fermions at $(N,2Q)=(24,62)$,
identified with 4/11.}
\label{fig4}
\end{figure}

\begin{figure}
\centerline{\psfig{figure=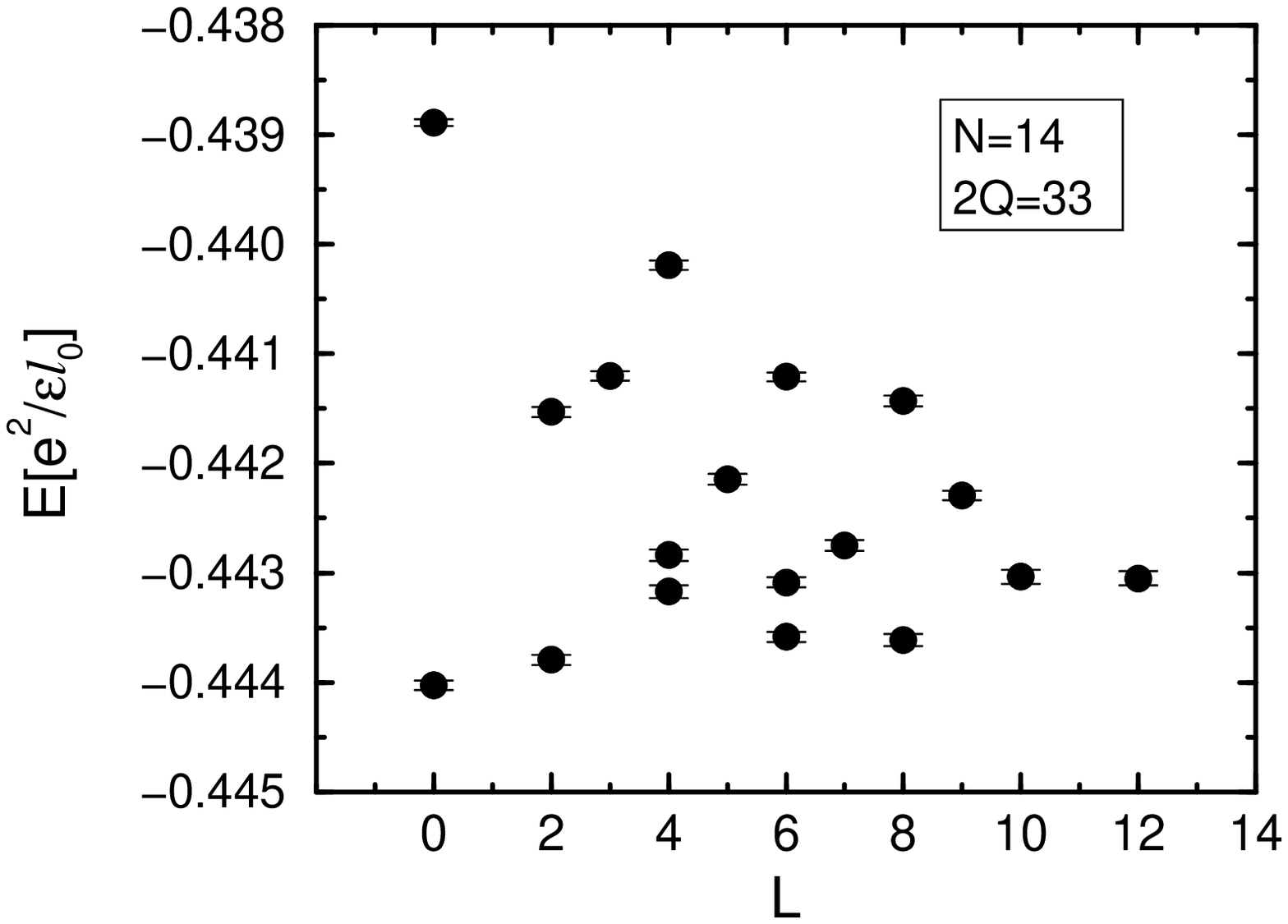,height=4.0in,angle=-90}}
\caption{The low energy spectrum of interacting composite fermions at $(N,2Q)=(14,33)$,
identified with 5/13.}
\label{fig5}
\end{figure}

\begin{figure}
\centerline{\psfig{figure=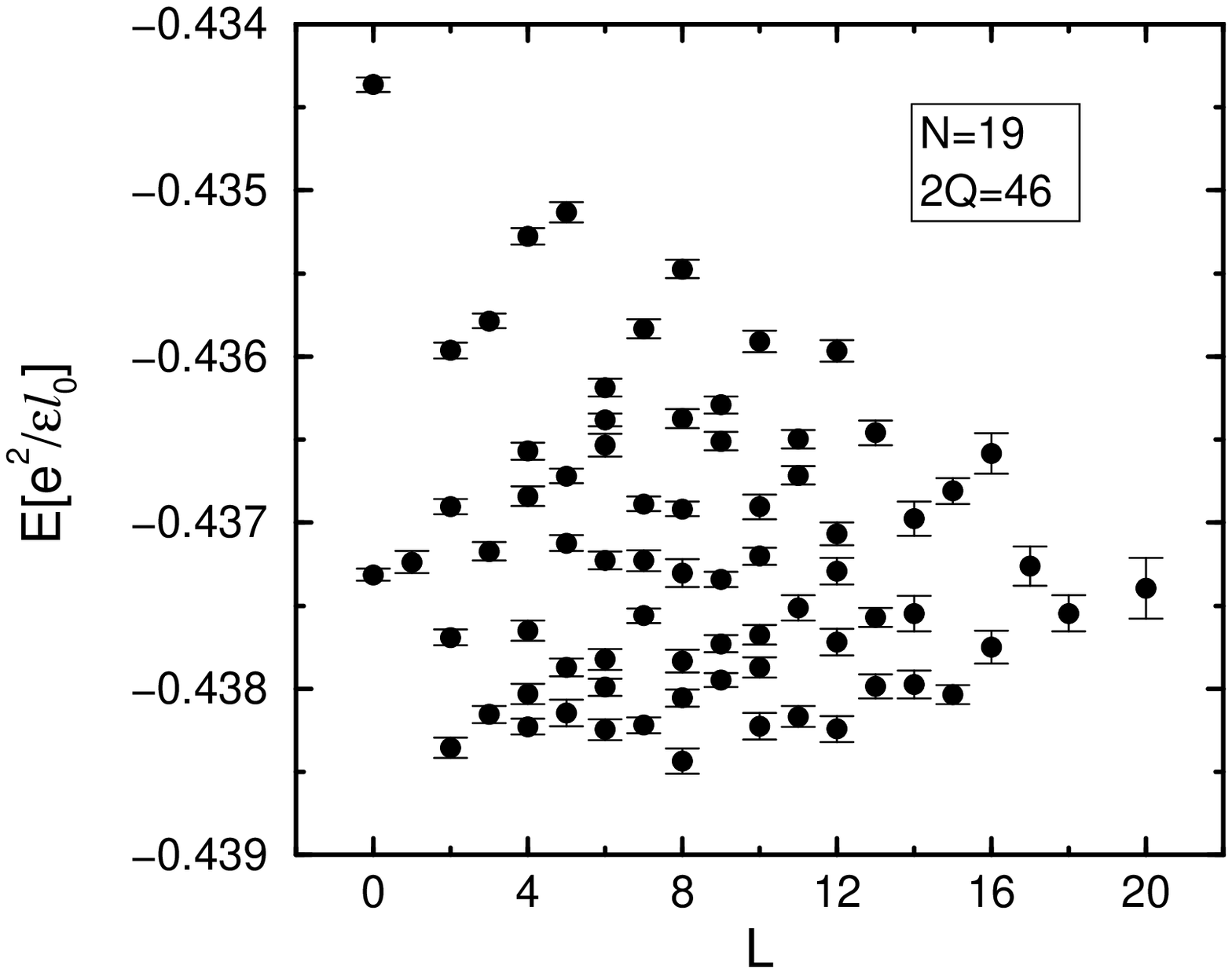,height=4.0in,angle=-90}}
\caption{The low energy spectrum of interacting composite fermions at $(N,2Q)=(19,46)$,
identified with 5/13.}
\label{fig6}
\end{figure}

\begin{figure}
\centerline{\psfig{figure=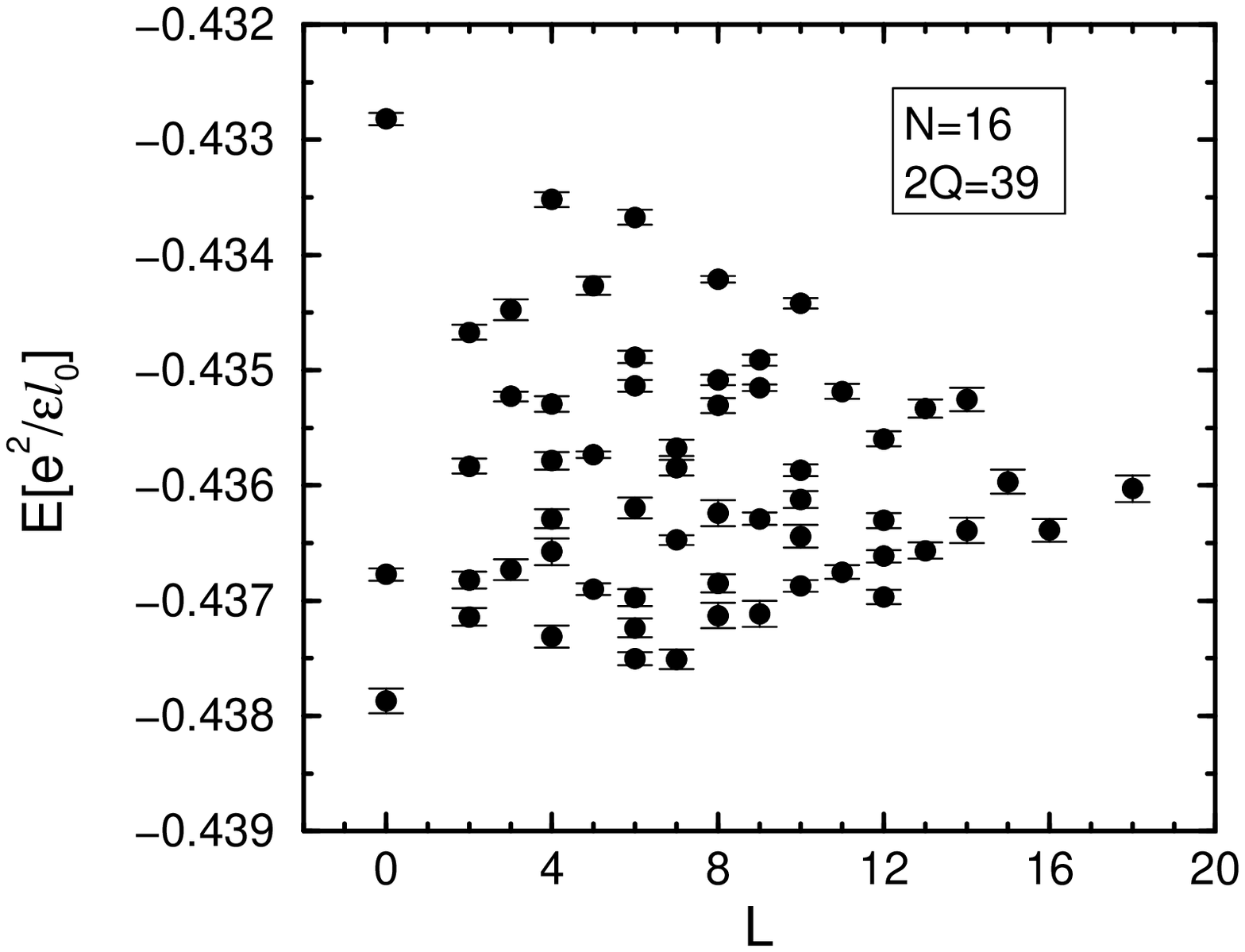,height=4.0in,angle=-90}}
\caption{The low energy spectrum of interacting composite fermions at $(N,2Q)=(16,39)$,
identified with 7/19.}
\label{fig7}
\end{figure}

\begin{figure}
\centerline{\psfig{figure=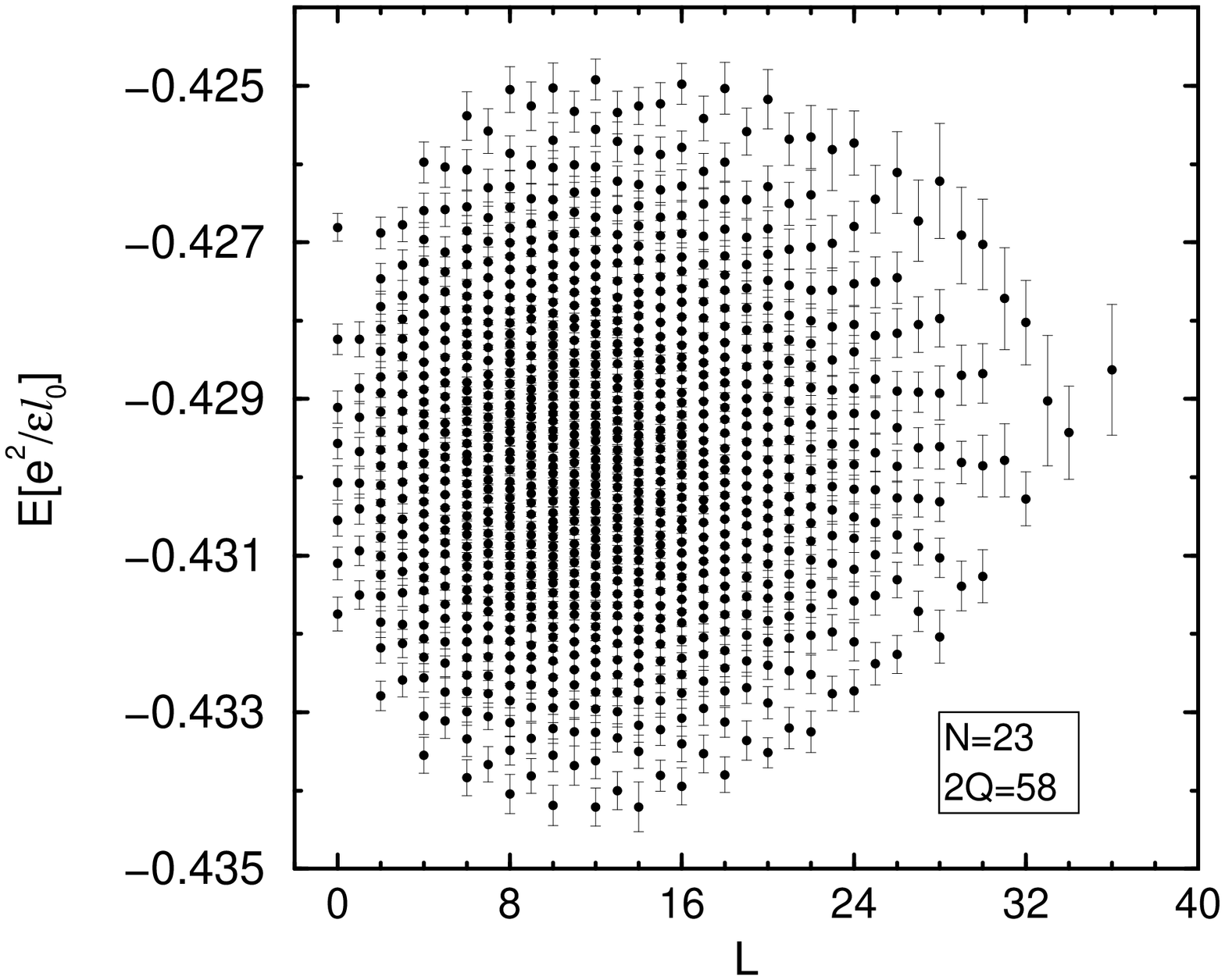,height=4.0in,angle=-90}}
\caption{The low energy spectrum of interacting composite fermions at $(N,2Q)=(23,58)$,
identified with 7/19.}
\label{fig8}
\end{figure}

\begin{figure}
\centerline{\psfig{figure=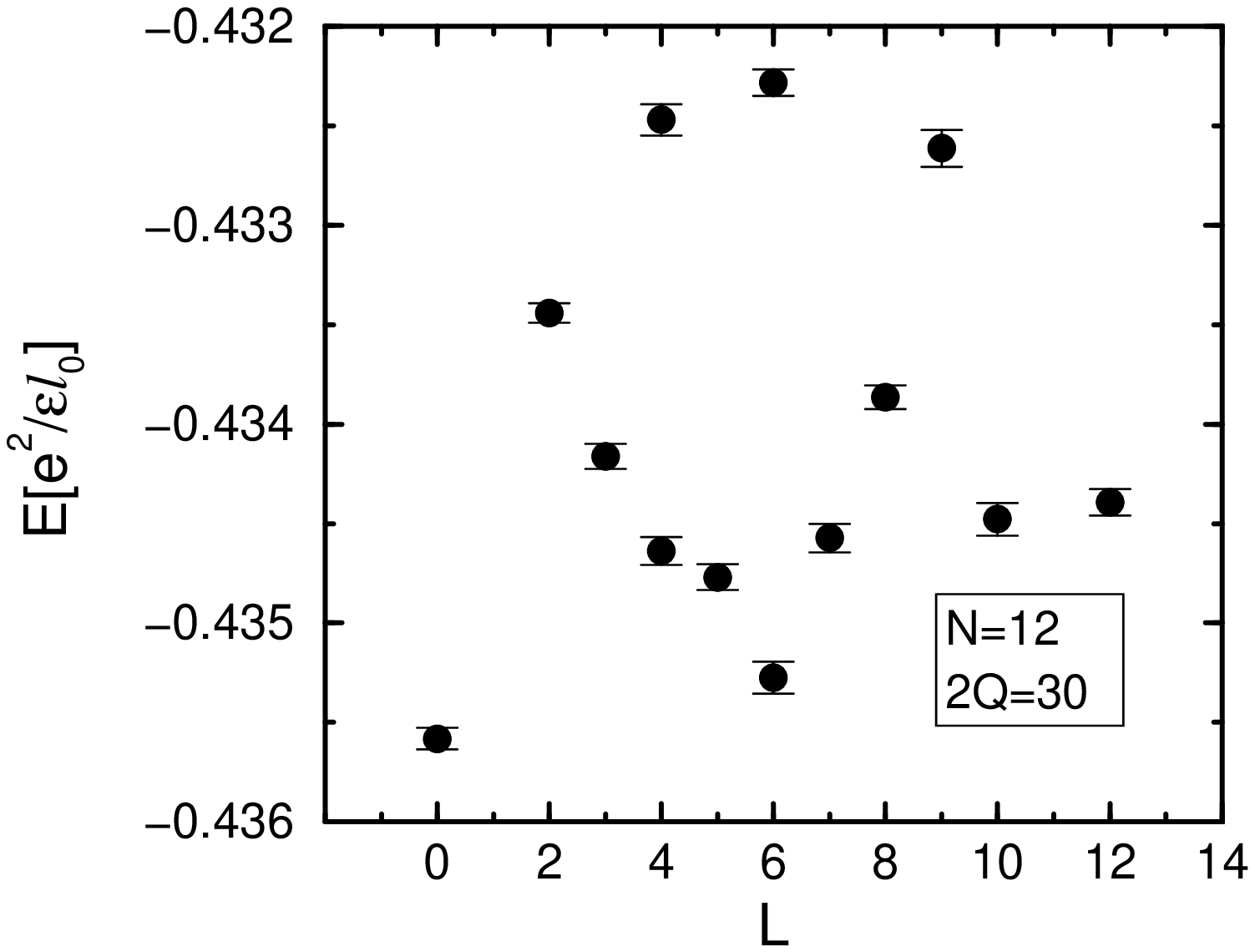,height=4.0in,angle=-90}}
\caption{The low energy spectrum of interacting composite fermions at $(N,2Q)=(12,30)$,
identified with 6/17.}
\label{fig9}
\end{figure}

\begin{figure}
\centerline{\psfig{figure=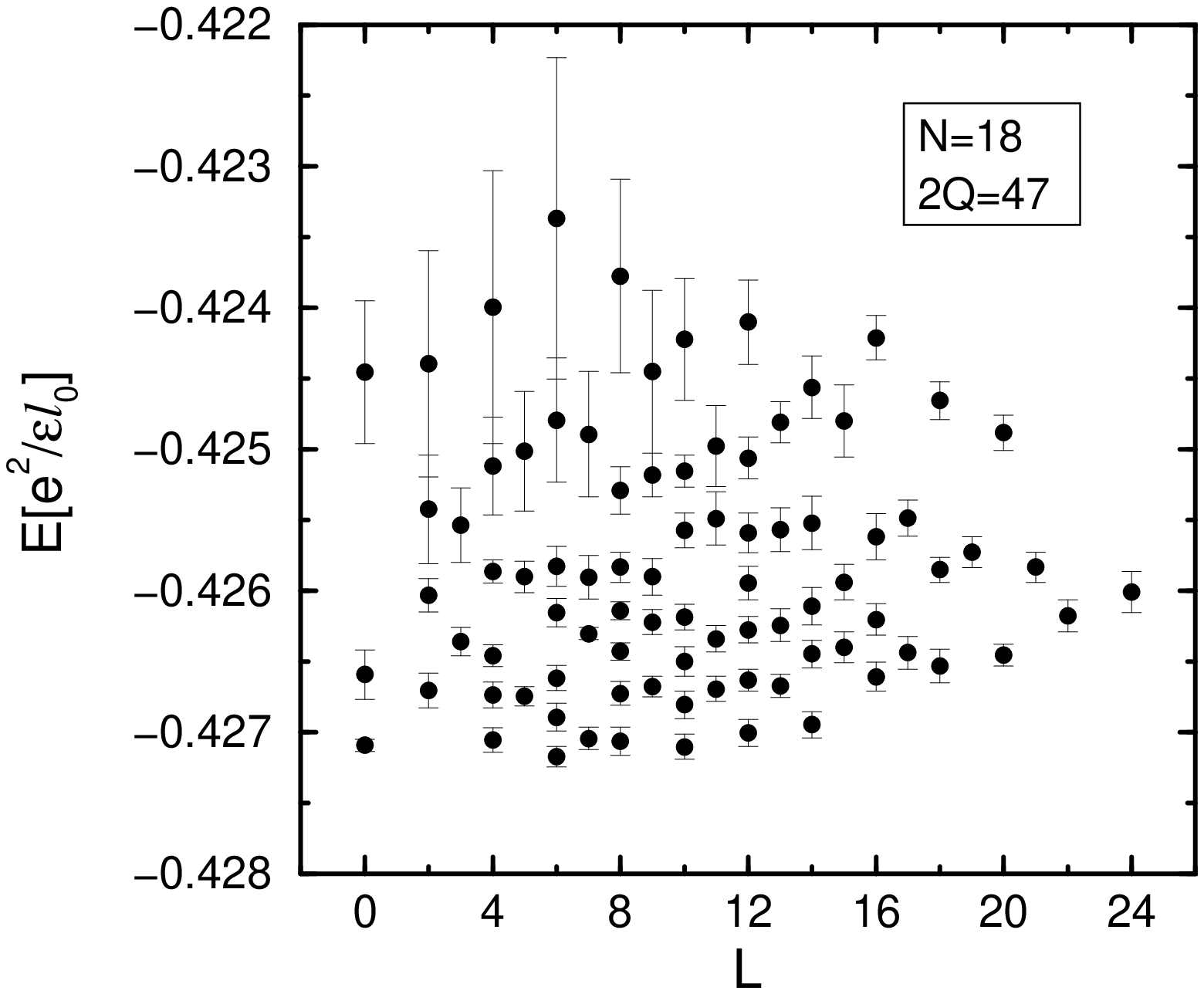,height=4.0in,angle=-90}}
\caption{The low energy spectrum of interacting composite fermions at $(N,2Q)=(18,47)$,
identified with 6/17.}
\label{fig10}
\end{figure}

\begin{figure}
\centerline{\psfig{figure=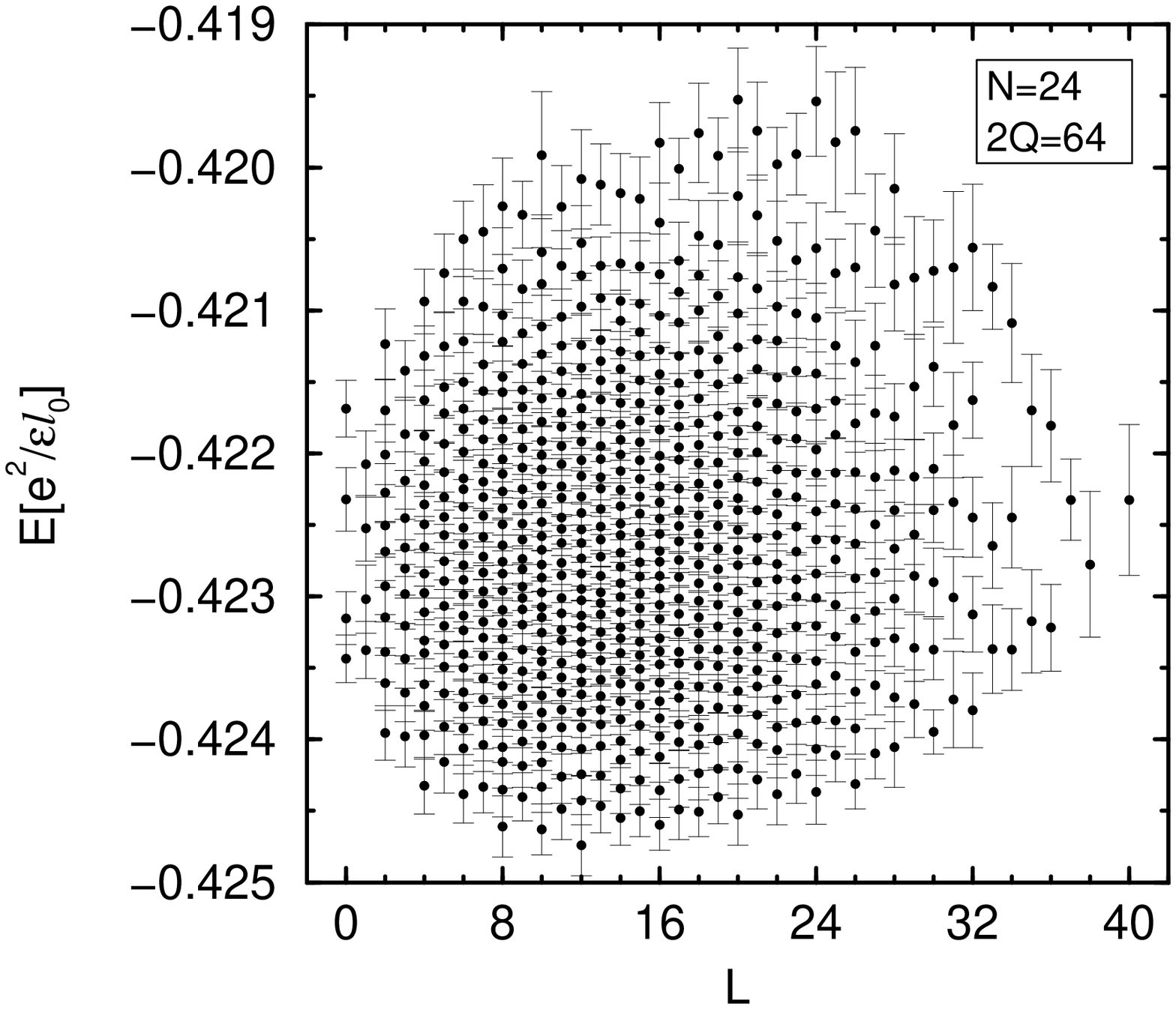,height=4.0in,angle=-90}}
\caption{The low energy spectrum of interacting composite fermions at $(N,2Q)=(24,62)$,
identified with 6/17.}
\label{fig11}
\end{figure}

We have obtained the low energy spectra at several values of $(N,2Q)$ which were identified 
earlier with systems at special filling factors 4/11, 5/13, 7/19, and 6/17.
Figs.~(\ref{fig1}), (\ref{fig2}), (\ref{fig3}), and (\ref{fig4}) show results for 4/11; 
Figs.~(\ref{fig5}), and (\ref{fig6}) for 5/13;  Figs.~(\ref{fig7}), and (\ref{fig8}) for 
7/19; and Figs.~(\ref{fig9}), (\ref{fig10}), and (\ref{fig11}) for 6/17.  
The energy spectrum for $(8,18)$ have been 
given in the past both from exact diagonalization and the composite fermion theory \cite{Dev,Book2,Quinn}. 
The accuracy of the energies is good enough to be able to state whether the ground state has $L=0$ 
or not.  The error increases with increasing $L$, which is a consequence of our use of 
an $L=0$ state for sampling; in principle, the error at large $L$ can be reduced by 
sampling with another reference state, but we do not see any need
for it for the question of interest in this work.

\begin{figure}
\centerline{\psfig{figure=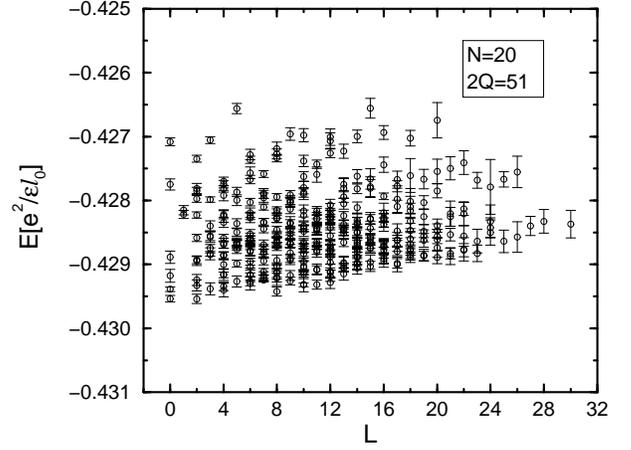,height=4.0in,angle=-90}}
\caption{The low energy spectrum of composite fermions at $(N,2Q)=(20,51)$ {\em prior} 
to orthogonalization.  The corresponding spectrum after orthogonalization is given in 
Fig.~\protect\ref{fig3}.
The spectra shown here and in Fig.~15 are not physically meaningful, but 
are given here only to show the effect of orthogonalization.}
\label{fig12}
\end{figure}

\begin{figure}
\centerline{\psfig{figure=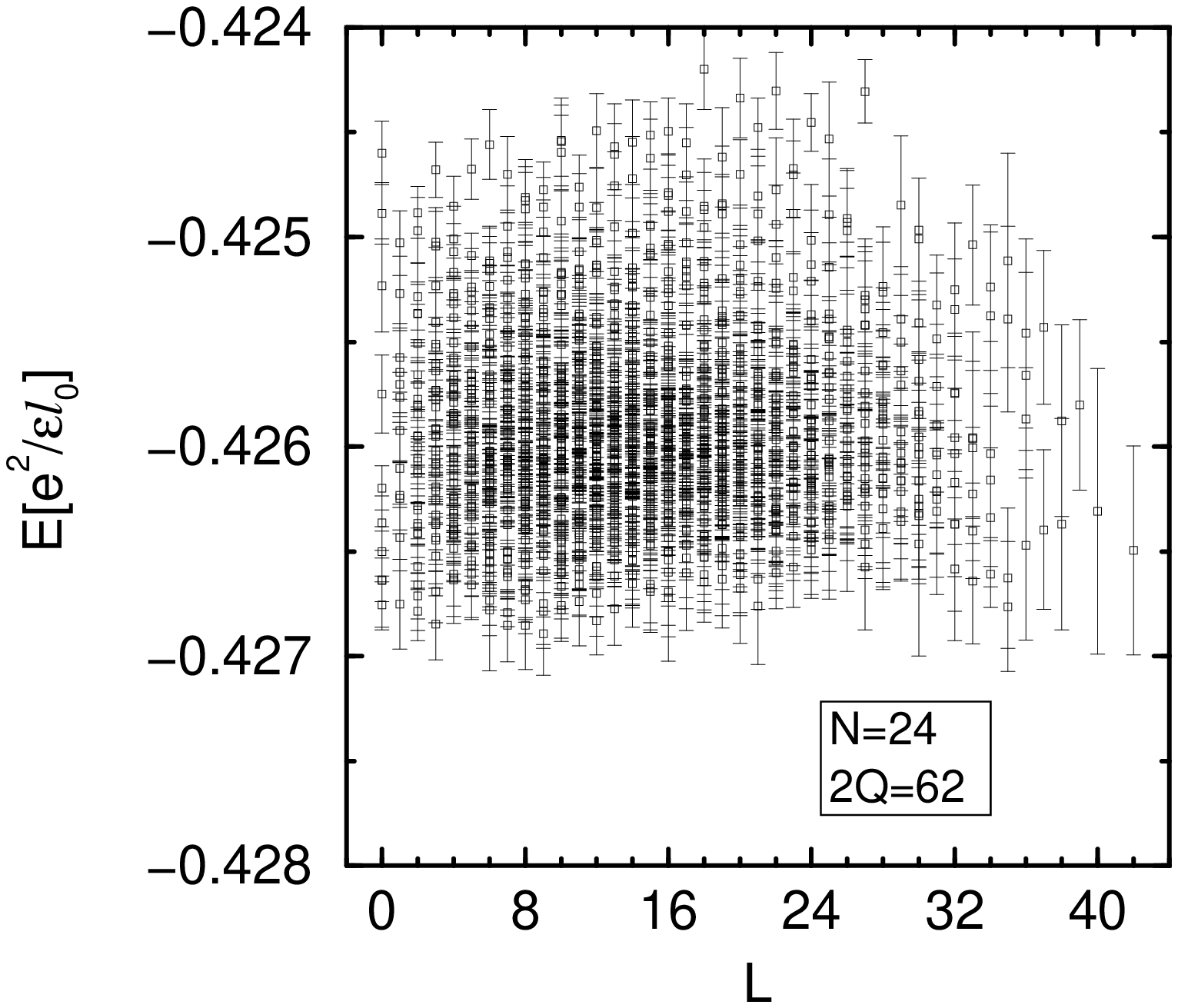,height=4.0in,angle=-90}}
\caption{The low energy spectrum of composite fermions at $(N,2Q)=(24,62)$ {\em prior} 
to orthogonalization.  It should be compared to the spectrum in Fig.~\protect\ref{fig4}.}
\label{fig13}
\end{figure}

The orthogonalization makes a qualitative difference in the spectrum.  Figs.~(\ref{fig12}) and 
(\ref{fig13}) show the energies of the basis states {\em prior} to orthogonalization, which 
ought to be compared to the spectra in Figs.~(\ref{fig3}) and (\ref{fig4}), respectively.

The results explicitly confirm that our method indeed gives very low energy states.  
For example, consider 4/11.  The ground state energy 
is approximately $-0.419 e^2/\epsilon l_0$ per particle, which compares very favorably with 
$-0.420 e^2/\epsilon l_0$, which is what one would get from a linear extrapolation between 
the energies \cite{JK}
at 1/3 ($-0.410 e^2/\epsilon l_0$) and 2/5 ($-0.433 e^2/\epsilon l_0$). 
Further, a large fraction of the states obtained here are within an energy band less than 
the  effective cyclotron gap at 1/3, which is on the order of 0.1 $e^2/\epsilon l_0$.
In other words, most of the states considered here lie within the 1/3 gap.
This justifies our approximation of neglecting the states that involve transitions between 
composite fermion Landau levels, which are expected to cost energy of order 0.1 $e^2/\epsilon l_0$.

Now we ask whether the spectra imply incompressibility.  A characteristic feature of 
{\em all} incompressible states known to date 
is that they have an $L=0$ ground state, which is separated by  
a gap from other states; in addition, there is also a well defined branch of neutral 
excitation.  The ground state is extremely well described as a filled Landau level state of 
composite fermions, and the excitation as a particle hole pair excitation of composite 
fermion \cite{Dev,Book2,Scarola}.
This is the case, with no exception, for for systems with $(Q,N)$ related by $Q=(1\pm 1/2n)N-(1\pm n/2)$,
which are identified with $\nu=n/(2n\pm 1)$.

None of the spectra have these standard features of an incompressible state.  Even in cases when 
the ground states has $L=0$, there is no well defined CF-exciton branch. 
The final spectrum at $\nu$ thus bears little resemblance to the spectrum at $\bar \nu$ that 
we started out with, implying that 
the interaction between composite fermions in the partially filled 
level is quite different from that of electrons at $\bar \nu$ in the lowest Landau level. 
The scenario described in the introduction for NQHE is thus not borne out.

The ground states at the flux values considered above 
are in general not incompressible uniform states with $L=0$.  
Consider 4/11 for example.
It has been known \cite {Dev,Book2,Quinn} that for $N=8$ particle system at $Q=9$, 
which is identified above with 4/11, the ground state does not have $L=0$.  However, 
the system is effectively very small here, with only $\bar N=3$ composite fermions 
in the second level,  and one may question if the conclusion will hold up as $N$ increases. 
We find that for $N=12$ and 20 the ground state has $L=0$, but for $N=16$ and 24 it does not.
This provides an illustration for why a study of several values of $N$ is crucial; 
if one had only studied $N=12$, that would have led one to exactly the opposite conclusion.
Another example is the 12 particle system at 6/17, which has an $L=0$ ground state
and might suggest FQHE at 6/17; however, the spectra for $N=18$ and 24 at 6/17 explicitly 
clarify that this is only a finite size effect that does not survive in the 
thermodynamic limit.

\begin{figure}
\centerline{\psfig{figure=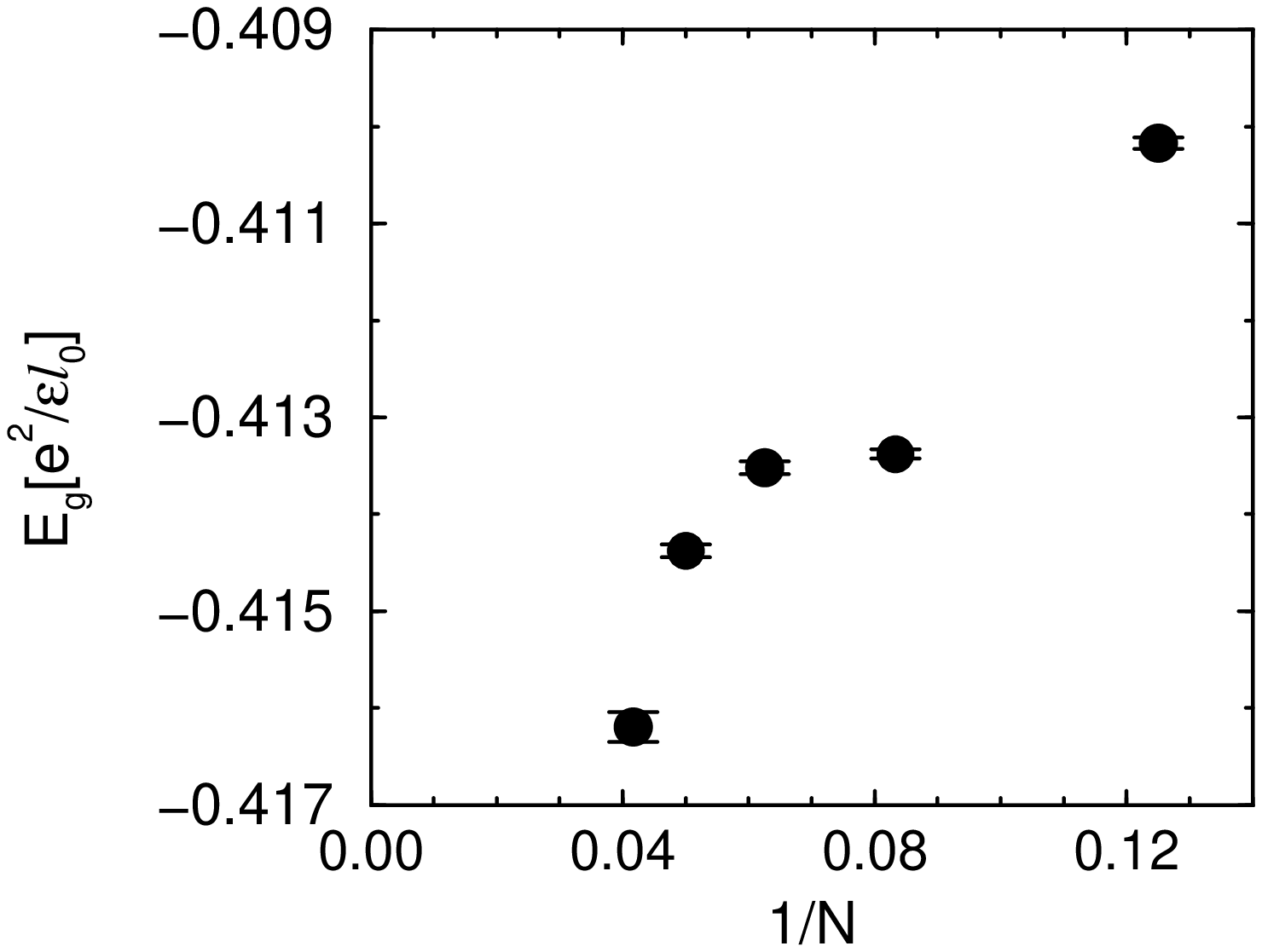,height=4.0in,angle=-90}}
\caption{The ground state energy per particle as a function of $1/N$ for 4/11.  The 
energies plotted here have been multiplied by the factor $\sqrt{\rho_N/\rho}$, where $\rho_N$ is 
the density of the finite system and
$\rho$ is the density in the thermodynamic limit, to account for the dependence of 
density on $N$.  It ought to be noted that the ground state does not have $L=0$ in general.}
\label{fig14}
\end{figure}

Our study implies that even though the ground state sometimes has 
$L=0$, it does {\em not} describe an  
incompressible state at any of the filling factors considered here.  
The following observations support this statement:

\begin{enumerate}

\item When the ground state has $L\neq 0$, the $L=0$ state is a rather highly excited state.

\item In the plot of the ground state energy with $1/N$, the energies of the $L=0$ ground
states are not particularly low. See Fig.~(\ref{fig14}) for 4/11.

\item The ground state at $\bar \nu$ generally produces an excited state
at $\nu$, even in the $L=0$ sector (prior to orthogonalization), which indicates a breakdown of 
analogy between $\bar \nu$ and $\nu$.  

\item The energy per particle is not a smooth function of $1/N$, but instead 
shows fluctuations characteristic of a compressible state. See Fig.~(\ref{fig14}). 

\item It is not inconceivable for a compressible
state to have $L=0$, because a uniform state avoids Hartree energy cost, but it would
be difficult to understand how an incompressible state might have $L\neq 0$.

\end{enumerate}

As a result, the physical mechanism, in which some of the composite fermions 
turn into higher order composite fermions and condense into Landau levels to exhibit QHE, 
does not appear to be relevant for fully spin polarized electrons.  
The interaction between composite fermions in the second level is thus not 
of a form to produce NQHE.
It must be stressed, however, that the results do {\em not} rule out NQHE through some 
hitherto unknown mechanism, which would imply a different relation between $2Q$ and $N$ for 
a given filling factor than the one assumed above.

In Ref.~\onlinecite{LSJ}, which also investigated some of the above fractions, it
was predicted that the ground state is a bubble crystal of composite fermions, 
similar to what was found in Hartree Fock studies of corresponding electronic states in 
higher Landau levels\cite{Koulakov}. Aside from several approximations, the 
results of that study were also predicated upon the quantitative validity of certain 
trial wave functions, making an independent confirmation worthwhile.
Unfortunately, the spherical geometry used
here is not convenient for the study of charge density wave states. 
We do note in passing that a crucial step in Ref.~\onlinecite{LSJ} was
to model the states at $\nu=n+\bar \nu$ in terms of fermions at $\bar \nu$ with an
effective two-body interaction.  A close similarity between our 
Fig.~(\ref{fig9}) and Fig.~(5) of Ref.~\onlinecite{Quinn}, obtained from the two-body interaction 
model [both for $(N,2Q)=(12,30)$] provides support to the validity of this model.

In the end, we note that the above calculations assume a strictly two dimensional 
system (with zero width), no disorder, and no mixing with higher electron Landau levels.  
These approximations represent an idealization of the actual experimental system.  It 
is expected that the idealized model, while 
not accurate {\em quantitatively}, should give the correct qualitative behavior.

It is a pleasure to acknowledge partial support by the National Science Foundation 
under Grant No. DMR-9986806.  We are grateful to Horst Stormer, Wei Pan, and Chia-Chen 
Chang for valuable
discussions, to David Weiss for suggesting the name "NQHE," to the   
High Performance Computing (HPC) group led by V. Agarwala, J. Holmes, and J. Nucciarone, 
at the Penn State University ASET (Academic Services and Emerging Technologies) for assistance
and computing time with the LION-XE cluster, and 
acknowledge NSF DGE-9987589 for computer support.

\pagebreak

\begin{table}[t]
\caption{The monopole strength $Q$ as a function of $N$ for the NQHE states 
studied in this work.  The quantities $\nu$ and $\nu^*$ are the the filling factors of electrons and 
composite fermions, with $\bar \nu$ being the CF filling factor in the partially filled level. 
$N$ is the total number of composite fermions, and $\bar N$ is the number of composite fermions 
in the partially filled level, at filling $\bar \nu$.  $Q$ and $q^*$ are the monopole strengths 
for the electrons and composite fermions.   
\label{tab1}}
\begin{center}
\begin{tabular}{|c|c|c|c|c|c|} 
$ \nu$         & $\nu^*=n+\bar \nu$         &  $N$  &  $\bar{N}$  &   $q^*$   &    $Q$  \\ \hline
$\frac{4}{11}$ & $1+\frac{1}{3}$ &   8   &   3         &  2        &   9   \\ \cline{3-6}
               &                 &   12  &   4         &  3.5      &  14.5  \\ \cline{3-6}
               &                 &   16  &   5         &  5      &  20  \\ \cline{3-6}
               &                 &   20  &   6         &  6.5      &  25.5  \\ \cline{3-6}
               &                 &   24  &   7         &  8      &  31  \\ \hline 
$\frac{5}{13}$ & $1+\frac{2}{3}$ &   14   &   6         &  3.5        &   16.5   \\ \cline{3-6}
               &                 &   19   & 8           &  5       & 23  \\ \hline
$\frac{7}{19}$ & $1+\frac{2}{5}$ &   16   &   6         &  4.5        &   19.5   \\ \cline{3-6}
               &                 &   23   & 8           &  7       & 29    \\ \hline
$\frac{6}{17}$ & $1+\frac{1}{5}$ &   12   &   3         &  4        &   15 \\ \cline{3-6}
               &                 &   18   &  4          & 6.5       & 23.5 \\ \cline{3-6}
               &                 &   24   &  5          &  9        & 32  \\ 
\end{tabular}
\end{center}
\end{table}


\pagebreak

\pagebreak

\begin{references}
\bibitem{warning}  We warn the reader that the assignment of filling factor to 
a finite system involves physical interpretation, and therefore must be treated 
with caution.  The relation between $N$ and $2Q$ at filling 
factor $\nu$ must have the form $N=2\nu Q + \alpha$, where $\alpha$ is of oder 
unity, so as to produce the correct filling factor in the limit 
$N\rightarrow \infty$ ($\lim_{N\rightarrow \infty} N/2Q \rightarrow \nu$), but the 
value of $\alpha$, which is crucial for finite 
systems, depends on the physics of the incompressible state. In this paper, the 
filling factor is attributed to a finite system assuming  
the familiar physics that has been successful in explaining 
the fractions $n/(2pn\pm 1)$.  A different origin of incompressibility could possibly 
imply a different relation between $N$ and $2Q$.
\end{references}
\end{document}